# Semantic annotation of requirements for automatic UML class diagram generation

Soumaya Amdouni[1], Wahiba Ben Abdessalem Karaa[2] and Sondes Bouabid[3]

[1] University of tunis High Institute of Management
Bouchoucha city, Bardo 2000, TUNISIA

[2] University of tunis High Institute of Management
Bouchoucha city, Bardo 2000, TUNISIA

[3] University of tunis High Institute of Management
Bouchoucha city, Bardo 2000, TUNISIA

**Abstract**

The increasing complexity of software engineering requires effective methods and tools to support requirements analysts' activities. While much of a company's knowledge can be found in text repositories, current content management systems have limited capabilities for structuring and interpreting documents. In this context, we propose a tool for transforming text documents describing users' requirements to an UML model. The presented tool uses Natural Language Processing (NLP) and semantic rules to generate an UML class diagram. The main contribution of our tool is to provide assistance to designers facilitating the transition from a textual description of user requirements to their UML diagrams based on GATE (General Architecture of Text) by formulating necessary rules that generate new semantic annotations.

## 1. Introduction

Increasing complexity of IS (information systems) and their quickly development prompted an increased interest in their study, in order to evaluate their performance in response to users' expectations. There is much and growing interest in software systems that can adapt to changes in their environment or their requirements in order to continue to fulfill their tasks. In fact, requirements specification is a fundamental activity in all process of software engineering. Many Researches [5] notice that many system failures can be attributed to a lack of clear and specific information requirements.

Knowledge requirements are formally defined and transferred from some knowledge source to a computer program. It has been argued that requirements study and knowledge acquisition are almost identical processes. Analysts can use several techniques necessary to extract relevant knowledge for software engineering. These knowledge define system expectations in terms of mission objectives environment, constraints, and measures of effectiveness and suitability. Thus, we need platforms and tools that enable the automation of activities involved in various life cycle phases of software engineering. These tools are very useful to extract functional and non-functional requirements from textual descriptions in order to develop graphic models of application screens, which will assist end-users to visualize how an application will look like after development. The aim of the work presented in this paper is to develop a tool that transforms a textual description to an UML class diagram. Our tool takes as input text data that represent textual user requirements descriptions. First, it identifies named entities (i.e., classes, properties and relationships between classes) and second it classifies them in a structured XML file.

The paper is organized into five sections. Section 2 reviews some related works. Section3 gives an overview of GATE API. Section 4 discusses our system and the final section presents a conclusion.

## 2. Related works

In the last years several efforts have been devoted by researchers in the Requirements Engineering community to the development of methodologies for supporting designers during requirements elicitation, modeling, and analysis.

However, these methodologies often lack tool support to facilitate their application in practice and encourage companies to adopt them.

The present work is in the context of engineering models such as MDA (Model Driven Architecture) which is a process based on the transformation of models: model to model, code to model, model to code, etc. It presents an experience in the application of requirements specifications expressed in natural language into structured specifications.

The proposed application having an input text data that represent user requirements identifies named entities (entities, properties and relationships between entities ....) to classify them in a structured XML file. Several researchers have tried to automate the generation of an UML diagram from a natural language specification.

Kaiya et al. [8] proposed a requirements analysis method which based on domain ontologies. However, this work does not support natural language processing, it allows the detection of incompleteness and inconsistency in requirements specifications, measurement of the value of the document, and prediction of requirements changes.

In [2] Christiansen et al. developed a system to transform use case diagram to class diagram Defnite Clause Grammars extended with Constraint Handling Rules. The grammar captures information about static world (classes and their relations) and subsequently the system generates the adequate class diagram. This work is very interesting but the problem that organization's requirements are not always modeled as use case diagram.

The work in [10] implemented a system named GeNLangUML (Generating Natural Language from UML) which generates English specifications from class diagrams. The authors translate UML version 1.5 class diagrams into natural language. This work was considered by most developers as an efficient solution for reducing the number of errors and verification and an early validation of the system but we need for all time to generate UML diagram from natural language. The system process is as follows:

- Grammatical labeling based on a dictionary wordnet to disambiguate the lexical structure of UML concepts.
- Sentences generation from the specification by checking attributes, operations and associations with reference to a grammar defining extraction rules.
- Checking if the generated sentences are semantically correct.
- Generating a structured document containing the natural specification of a natural class diagram.

Hermida et al. [7] proposed a method which adapts UML class diagrams to build domain ontologies. They describe the process and the functionalities of the tool that they have developed for supporting this process. The authors have chosen a use case in the pharmacotherapeutic domain. The authors present a good approach however it is specific to a well defined area (pharmacotherapeutic).

In [6] authors proposed a tool NT2OD which derives an initial object diagram from textual use case descriptions using natural language processing (NLP) and ontology learning techniques. NT2OD consists of creating a parse tree for the sentence, identifying objects and relations and generating the object diagram.

In our work we propose a CASE tool (Computer-aided software engineering). We extract information from users' requirements to generate class diagram taking in account existing approaches. We propose a design tool which extracts UML concepts and generate UML class diagram according to different concepts (class, association, attribute). The idea is to use GATE API[1] and we extended it by new JAPE rules to extract semantic information from user requirements.

## 3. GATE overview

GATE "General Architecture for Text Engineering" is developed by the Natural Language Processing Research Group [2] at the University of Sheffield [3]. GATE is a framework and graphical development environment, which enables users to develop and deploy language engineering components and resources in a robust fashion [4]. GATE contains different modules to process text documents. GATE supports a variety of formats (doc, pdf, xml, html, rtf, email...) and multilingual data processing using Unicode as its default text encoding.

In the present work we use the information extraction tool **ANNIE plugin** (**A N**early-**N**ew **IE** system) (Fig. 1). It contains Tokeniser, Gazetteer (system of lexicons), Pos Tagger, Sentence Splitter, Named Entity Transducer, and OrthoMatcher.

- Tokeniser: this component identifies various symbols in text documents (punctuation, numbers, symbols and different types). It applies basic rules to input text to identify textual objects.
- Gazetteer: gazetteer component creates annotation to offer information about entities (persons, organizations...) using lookup lists.
- POS Tagger: this component produces a tag to each word or symbol.
- Sentence splitter: sentence splitter identifies and annotates the beginning and the end of each sentence.

---

[1] http://gate.ac.uk/
[2] http://nlp.shef.ac.uk/
[3] http://www.shef.ac.uk/

- Named Entity Transducer**:** the NE transducer applies JAPE rules to input text to generate new annotations [1].

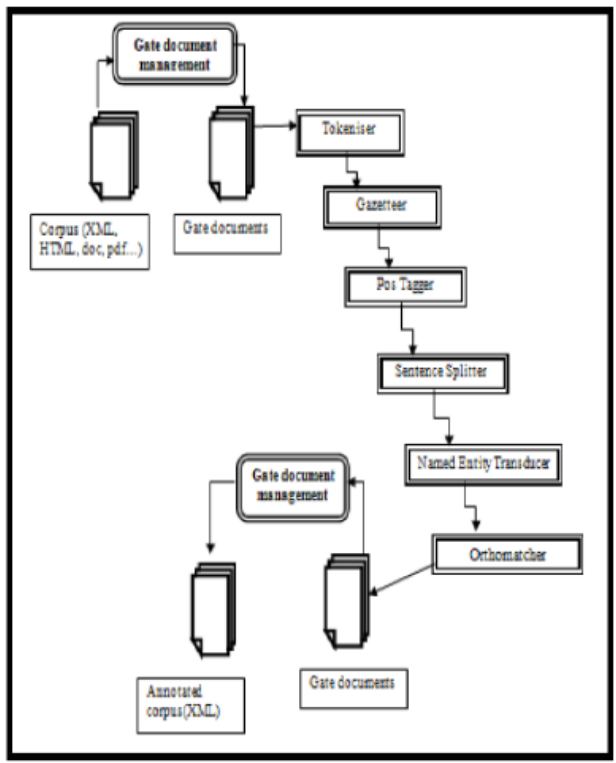

Fig. 1  ANNIE components in GATE.

## 4. System description

Our system is a document analysis and annotation framework that uses efficient methods and tools adopted from markup domain. The approach discriminates between domains of the annotation process and hence allows an easy adaptation to different applications.

In fact, it uses GATE API and especially the following components: sentence splitter, pos tagger, gazetteer, named entity transducer. The entity recognition is the most interesting task for this reason we extended ANNIE tool with additional rules and additional lists to enhance entities' extraction. The following figure (fig. 2) describes the process we have proposed for the extraction UML concepts in order to generate an UML class diagram.

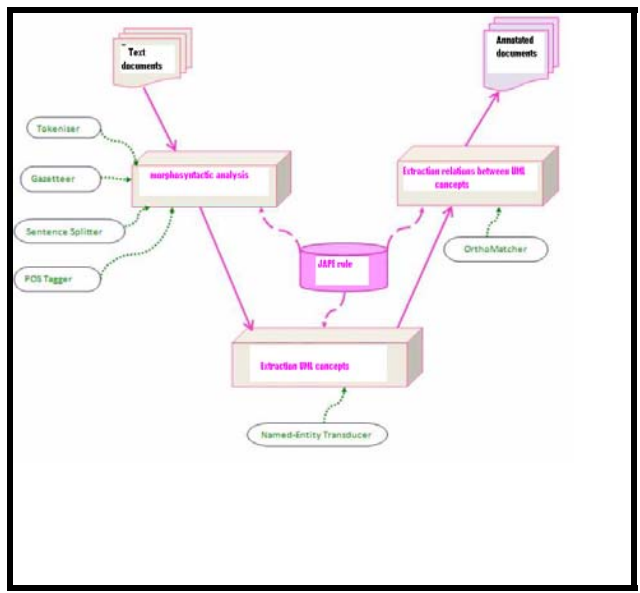

Fig. 2. System architecture.

### 4.1 Morphological analysis

The first phase of our system consists in morphosyntactic analysis of users' requirements. The tool parses an input document according to a predefined grammar. The produced parse tree consists of structures such as paragraph, sentence, and token. In this step we use sentence splitter and Tokeniser component to extract sentences and basic linguistic entities. Then, we used Pos Tagger to associate with each word (token) grammatical category and to distinguish the morphology of various entities. For example below, the tagger identifies a verb (i.e., passe), two nouns (i.e., client, commande), an, and two prepositions (i.e., le, une ).

**Le (PRP) client (NN) passe (VB) une (PRP) commande (NN).**

### 4.2 Semantic Extraction of UML concepts

The second phase is extraction of UML concepts. The system is based in the results generated by morphosyntactic analysis stage and uses the Named Entity Transducer component to perform the operation for extracting named entities (classes, attributes and associations) referring to new JAPE rules and Gazetteer lists.

JAPE rule (Java Annotation Patterns Engine) a variant adapted to the Java programming language consists in files containing a set of rules [3]. Gazetteer lists are lookup lists with one entry per line containing names of people large organizations, months of the year, days of the week, numbers, etc [10].
In class diagram usually have the following format:

**Noun+verb+Noun**

The example below demonstrates two classes (le client, une commande), and an association (passe).

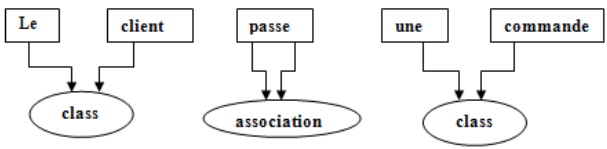

Figure 3 (fig. 3) describes a Jape rule called "Class" allowing the recognition of classes running a well-defined set of instructions. The actual treatment begins from line 7 of this figure by testing whether the word (token) under analysis belongs to a gazetteer list already defined. If this test is checked, the word in question will be annotated "Class" or there will be a passage to the following instructions from line 16.

```
1.  Phase:IdentifClasse
2.  Input:  Lookup Token
3.  Options: control =appelt
4.  Rule: Classe
5.  Priority: 20
6.  (
7.  {Lookup.majorType ==ClassMinuscule }
8.  |
9.  {Lookup.majorType==ClassJustPremierMaj}
10. |
11. {Lookup.majorType==ClassPremierMaj}
12. |
13. {Lookup.majorType==ClassMajuscule}
14. |
15. (
16. {Token.kind==word, Token.category==NNP}
17. {Token.kind==word, Token.category==V}
18. )
19. ): label
20. -->
21. :label.classe = {rule= Classe}
```

Fig. 3. JAPE rule for extracting UML class.

To extract association concept we use Jape rule illustrated in Figure 4 (fig. 4). If the token belong to gazetteer lists (lines 7, 9, 11, 13), it will be annotated as association otherwise the instructions from line 16 will be executed: if the token belongs to the class list, the second token is a "verb", and that the third word belongs to the list "Class", then the second word (token) will be annotated as an association.

```
1.  Phase:IdentifAssociation
2.  Input:  Lookup Token
3.  Options: control = appelt
4.  Rule: Association
5.  Priority: 20
6.  (
7.  {Lookup.majorType == AssociationMinuscule }
8.  |
9.  {Lookup.majorType== AssociationTTMajuscule}
10. |
11. {Lookup.majorType== AssociationMajJustDéb}
12. |
13. {Lookup.majorType== AssociationMajDéb }
14. |
15. (
16. {Lookup.majorType == class }
17. {Token.kind==word, Token.category==V}
18. {Lookup.majorType == class }
19. )
20. ): label
21. -->
22. : label.Association={rule=Association}
```

Fig. 4 JAPE rule for extracting association.

In addition, we execute instructions in figure 5 (fig. 5) to extract attribute. This rule is running as precedent ones (JAPE rule extracting class, and JAPE rule extracting association). If the token fits in attribute lists so it will have an attribute annotation. Else if the token is a name following by a verb and another name not belonging in the class list the latter is identified as an attribute.

```
1.  Phase:IdentifAttribut
2.  Input:  Lookup Token
3.  Options: control = appelt

4.  Rule: Attribut
5.  Priority: 20
6.  (
7.  {Lookup.majorType ==attribut Mini}
8.  |
9.  {Lookup.majorType==attribut TT MAJ}
10. |
11. {Lookup.majorType==Attribut Maj Déb Chak Mot}
12. |
13. {Lookup.majorType==attribut justDébutMaj}
14. |
15. (
16. {Token.kind==word, Token.category==NNP}
17. {Token.kind==word, Token.category==V}
18. {Token.kind==word,
     Token.category==NNP},{Lookup.majorType != class }
19. )
20. ): label
21. -->
22. :label.Attribut = {rule= Attribut}
```

Fig. 5  JAPE rule for extracting attribute.

In this step, we propose a graphical representation of JAPE rules set used by all modules of ANNIE components that we have integrated in our application (fig. 6).

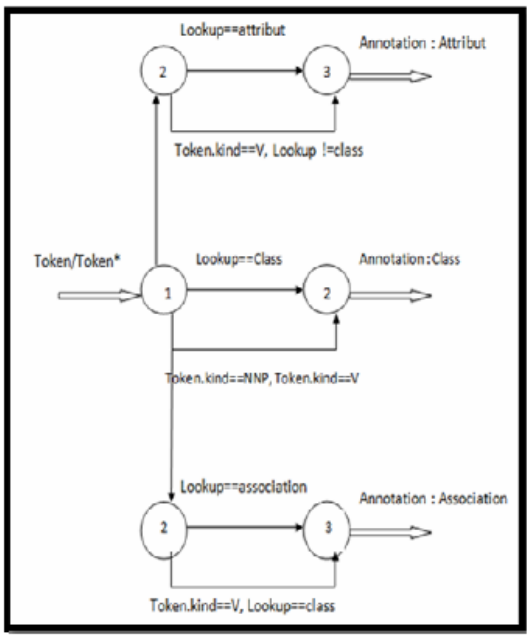

Fig. 6. General transducer

Figure 6 (fig. 6) illustrates a transducer describing extended JAPE grammar used in the context of our proposal: semantic extraction of UML concepts in order to create the corresponding UML class diagram.

### 4.3 Extraction relations between UML concepts

The third phase allows organizing relations between the entities (UML concepts) and gives not defined entities the corresponding annotation based on relations between the named entities that already exist. This phase presents a coreference resolution which is executed by Orthomatcher component. The tasks of recognizing relations are more challenging. The tool matches and annotates complex relations using annotations rules.

### 4.3 Test phase

In this phase, we have formed a corpus of users' requirements in different areas. Then, we have tested our system on this corpus. We applied GATE which generates an XML file containing all semantic tags. We clean the file by removing unnecessary tags like <sentence>, <token>… Figure 7 (fig. 7) shows an example of output GATE file. Our tool is robust and efficient and the error rate is very low, except that case studies are very complicated.

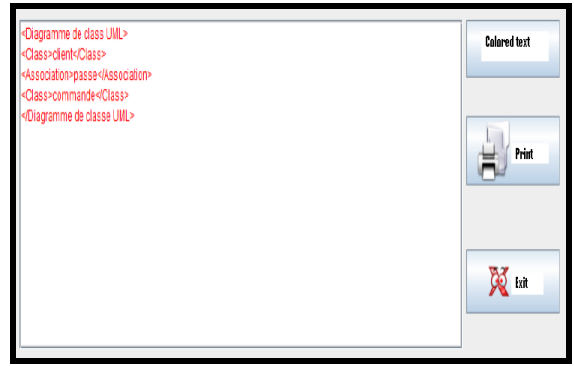

Fig. 7. GATE output file

### 5. Conclusion

Documents are central to Knowledge Management, but intelligent documents, created by semantic annotation, would bring the advantages of semantic search and interoperability. These benefits, however, come at the cost of increased authoring effort. Our system consists in semantic annotation of users' requirements based in GATE API. In fact, we have followed JAPE rules and Gazetteer lists elaboration to identify classes, associations and attributes in class diagrams. We assume that a chart generation of our XML file will be useful to ensure good readability for the designer. This work is already underway.

### Acknowledgments

The authors would like to thank GATE users for their disponibilities and their helps.

**Soumaya Amdouni** is a PhD candidate at the University of tunis High Institute of management. She received her master's degree in computer science in April 2010 from High Institute of Management. Her research interest includes natural language processing, semantic annotation, and web service.

**Wahiba Ben Abdessalem** is an assistant professor in the Department of Computer and Information Science at university of Tunis High Institute of Management. She received the Master Degree in 1992 from Paris III, New Sorbonne, France, and PhD, from Paris 7 Jussieu France in 1997. Her research interest includes Modelling Information System, Natural language processing, document annotation, information retrieval, text mining. She is a member of program committee of several International Conferences: ICCA'2010, ICCA'2011, RFIW 2011, and a member of the Editorial Board of the International Journal of Managing Information Technology (IJMIT).

**Sondes Bouabid** is a student in master in computer science at Paris dauphine. She obtained her degree in June 2010 from University of Tunis High Institute of Management. Her research interests are: textmining, and information system.